\documentclass[showpacs,superscriptaddress,twocolumn,pra]{revtex4}
\usepackage{amsmath,amssymb}
\usepackage{graphicx}

\begin{document}

\title{Light induced effective magnetic fields for ultra-cold atoms in planar
geometries}

\date{\today}

\author{G. Juzeli\={u}nas}
\affiliation{Institute of Theoretical Physics and Astronomy of 
Vilnius University,\\
A. Go\v{s}tauto 12, 01108 Vilnius, Lithuania}
\author{J. Ruseckas}
\affiliation{Institute of Theoretical Physics and Astronomy of 
Vilnius University,\\
A. Go\v{s}tauto 12, 01108 Vilnius, Lithuania}
\affiliation{Fachbereich Physik, Technische Universit\"at Kaiserslautern,\\
D-67663 Kaiserslautern, Germany}
\author{P. \"Ohberg}
\affiliation{Department of Physics, University of Strathclyde,\\
Glasgow G4 0NG, Scotland}
\author{M. Fleischhauer}
\affiliation{Fachbereich Physik, Technische Universit\"at Kaiserslautern,\\
D-67663 Kaiserslautern, Germany}

\begin{abstract}
We propose a scheme to create an effective magnetic field for ultra-cold atoms
in a planar geometry. The set-up allows the experimental study of classical and
quantum Hall effects in close analogy to solid-state systems including the
possibility of finite currents. The present scheme is an extention of the
proposal in Phys.~Rev.~Lett.\ \textbf{93}, 033602 (2004) where the effective
magnetic field is now induced for three-level $\Lambda$-type atoms by two
counterpropagating laser beams with shifted spatial profiles.  Under conditions
of electromagentically induced transparency the atom-light interaction has a
space dependent dark state, and the adiabatic center of mass motion of atoms in
this state experiences effective vector and scalar potentials.  The associated
magnetic field is oriented perpendicular to the propagation direction of the
laser beams.  The field strength achievable is one flux quantum over an area
given by the transverse beam separation and the laser wavelength. For a
sufficiently dilute gas the field is strong enough to reach the lowest Landau
level regime. 
\end{abstract}


\maketitle

One of the most fascinating subjects at the interface between ultra-cold atoms
and solid-state systems is the possibility to experimentally study
strong-correlation phenomena with the precision and the large degree of
variability provided by atomic physics. For example interacting Bose-Einstein
condensates (BEC) or degenerate Fermi gases in rotating two-dimensional traps
are studied in several laboratories with the goal to observe quantum-Hall like
effects \cite{bretin04,schweikhard04,baranov05}. The trap rotation provides an
effective magnetic field for the electrically neutral atoms. However in order to
reach the fractional quantum Hall regime it is necessary to rotate the trap
close to the critical frequency. Furthermore the atom density needs to be low
enough such that the number of magnetic flux quanta approaches the number of
atoms, which is an experimental challenge. Besides experimental difficulties
this approach has some conceptual drawbacks: It is limited to rotational
symmetric set-ups and does not allow to study transport phenomena, i.e. the
effect of magnetic fields to a finite particle current.
 
In \cite{juzeliunas04,juzeliunas05} we have suggested an alternative method
based on light-induced gauge potentials for atoms with a space-depended dark
state. A dark state is created if three-level $\Lambda$-type atoms interact with
two laser fields under conditions of electromagnetically induced transparency
(EIT) \cite{arimondo96,harris97,matsko01,lukin03,fleischhauer05}. If the dark
state is space dependent, a vector gauge potential arises for the adiabatic
center-of-mass motion \cite{dum96}. As shown in \cite{juzeliunas04,juzeliunas05}
the vector potential is associated with a nonvanishing magnetic field, if at
least one of the two light beams has a vortex i.e. an orbital angular momentum
(OAM).  Yet the use of vortex light beams has similar drawbacks as the trap
rotation regarding the spatial symmetry and transport phenomena. 

\begin{figure}[hbt]
\begin{center}
\includegraphics[width=8 cm]{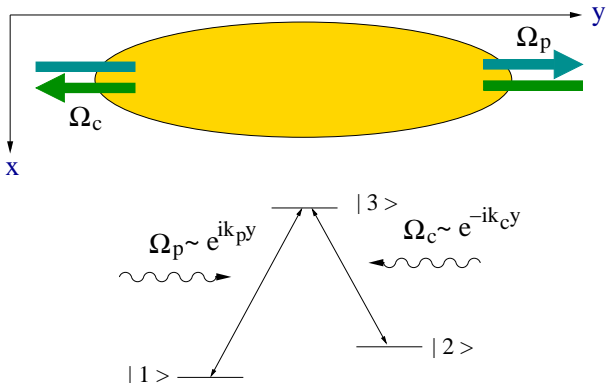}
\end{center}
\caption{(Color online) (top) Schematic representation of set-up for
light-induced effective magnetic field: Two counterpropagating and overlapping
laser beams interact with a cloud of cold atoms.  (bottom) The level scheme for
the $\Lambda$-type atoms interacting with the resonant probe and control beams
characterized by Rabi frequencies $\Omega_p$ and $\Omega_c$.}
\label{fig:1}
\end{figure}

We here propose a variation of this scheme which is free of the above mentioned
limitations. The scheme, shown in Fig.~\ref{fig:1} once again involves two laser
beams interacting with three-level atoms in the EIT configuration. Yet we are no
longer dealing with light beams posessing an OAM with respect to their
propagation axis.  As we will show lateron a nonvanishing magnetic field
requires only a \textit{relative} OAM between the two light beams.  This can be
achieved by two counterpropagating and overlapping laser beams with shifted
spatial profiles. In this case an effective magnetic field appears perpendicular
to the propagation direction and to the gradient of the relative intensity of
the light beams.  This configuration allows a planar geometry and a nonvanishing
flow of atoms, e.g. an atomic BEC moving along an atomic waveguide
\cite{Folman02Atom-Chip}. 

Let us consider an ensemble of cold three-level atoms with lower levels
$|1\rangle$ and $|2\rangle$ and electronically excited state $|3\rangle$.  The
atoms interact with two resonant laser beams in the EIT configuration, see
Fig.~\ref{fig:1}.  The first beam (to be referred to as the control beam) has a
frequency $\omega _{c}$, a wave-vector $\mathbf{k}_{c}$, and induces the atomic
transitions $\left| 2\right\rangle \rightarrow \left| 3\right\rangle$ with Rabi
frequency $\Omega _{c}\equiv \mu_{32}E_{c}/2$, where $E_{c}$ is the electric
field strength and $\mu _{32}$ is the transition  dipole moment. The second
(probe) beam with frequency $\omega _{p}$, wave-vector $\mathbf{k}_{p}$ causes
the transition $\left| 1\right\rangle \rightarrow \left| 3\right\rangle $ with a
Rabi frequency $\Omega _{p}\equiv \mu _{31}E_{p}/2$.  The two laser beams keep
the atoms in their dark state
\cite{arimondo96,harris97,matsko01,lukin03,fleischhauer05}:
\begin{equation}
|D\rangle=|1\rangle \cos\theta -|2\rangle \sin\theta \exp(iS) \sim |1\rangle
-\zeta|2\rangle,
\label{dark-state}
\end{equation}
where $\zeta =\Omega _{p}/\Omega _{c}= \left| \zeta\right| \mathrm{e}^{iS}
\equiv \tan\theta \, \mathrm{e}^{iS} $ is the ratio between the Rabi frequencies
of the probe and control fields, $S$ is their relative phase, and $\theta$ is
the mixing angle between the states $|1\rangle$ and  $|2\rangle$ in the atomic
dark state $|D\rangle$.

The dark state depends on the atomic position through the
$\mathbf{r}$-dependence of the Rabi frequencies $\Omega_p(\mathbf{r})$ and
$\Omega_c(\mathbf{r})$, so  an effective vector potential (generally known as
the Berry connection \cite{jackiw88,sun90}) appears in the adiabatic equation of
motion for the atomic center of mass.  The effective vector and trapping
potentials governing the translational motion of the dark-state atoms read
\cite{juzeliunas05}:
\begin{equation}
\mathbf{A}_{\mathrm{eff}}= -\hbar\frac{|\zeta|^2}{1 +|\zeta^2|}\mathbf{\nabla}S
= -\hbar\sin^2\theta\, {\mathbf\nabla} S
\label{A-eff-rez}
\end{equation}
and 
\begin{equation}
V_{\mathrm{eff}}(\mathbf{r}) = V_{\mathrm{ext}}(\mathbf{r})+\frac{\hbar^2}{
2m}\frac{|\zeta|^2(\nabla S)^2+(\nabla|\zeta|)^2}{\left(1+|\zeta|^2\right)^2},
\label{V-eff-rez}
\end{equation}
where
\begin{equation}
V_{\mathrm{ext}}(\mathbf{r})=\frac{V_1(\mathbf{r})
+|\zeta|^2(V_2(\mathbf{r})+\hbar \omega_{21})}
{1+|\zeta|^2}
\label{vext}
\end{equation}
is the external potential for the dark-state atoms, $V_j(\mathbf{r})$ is the
trapping potential for an atom in the internal state $j$, and $\omega_{21}
=\omega_{2}-\omega _{1}+\omega_{c}-\omega _{p}$  is the frequency of the two
photon detuning.

One easily recognizes that the vector gauge potential
$\mathbf{A}_{\mathrm{eff}}$ yields a non-vanishing magnetic field only if the
gradients of the relative intensity and the relative phase are both non-zero and
not parallel to each other:
\begin{equation}
\mathbf{B}_{\mathrm{eff}}\equiv \nabla \times \mathbf{A}_{\mathrm{eff}} = -\hbar
\nabla(\sin^2\theta)\, \times\, \nabla S.
\label{eq:Beff-general}
\end{equation}
This equation has a very intuitive interpretation: $\nabla(\sin^2\theta)$ is a
vector that connects the ``center of mass'' of the two light beams, $\nabla S$
is proportional to the vector of the relative momentum of the two light beams.
Thus a nonvanishing $\mathbf{B}_{\mathrm{eff}}$ requires a \textit{relative
orbital angular momentum} of the two light beams. As discussed in
\cite{juzeliunas04,juzeliunas05} this is the case e.g. for light beams with a
vortex. 

Here we consider however a different scenario. We suggest to use two
counterpropagating light beams of finite diameter with an axis offset: $\Omega
_{p}=\Omega_{p}^{(0)}e^{ik_{p}y}$ and $\Omega _{c}=\Omega
_{c}^{(0)}e^{-ik_{c}y}$, where $\Omega _{p}^{(0)}$ and $\Omega _{c}^{(0)}$ are
real amplitudes with shifted transverse profiles. The beams possess a relative
orbital angular momentum similarly to two point particles with constant momenta
passing each other at some finite distance. In such a situation the phase of the
ratio $\zeta =\Omega_{p}/\Omega _{c}$ is given by: 
\begin{equation}
S=ky, \quad  k=k_{p}+k_{c},
\end{equation}
so that $\nabla S= k \hat{\mathbf{e}}_{y}$ where $\hat{\mathbf{e}}_{y}$ is a
unit Cartesian vector.

The spatial dependence of the intensity ratio $|\zeta |^{2}=\left| \Omega
_{p}/\Omega _{c}\right| ^{2}$ is determined by the spatial profiles of both
$\left| \Omega _{p}\right| ^{2}$ and $\left| \Omega _{p}\right| ^{2}$. Since the
control and probe beams counterpropagate along the the $y$-axis, their
intensities depends weakly on $y$. Furthermore we shall disregard the
z-dependence of the intensity ratio $|\zeta |^{2}$.  This is legitimate, for
instance, if the atomic motion is confined to the xy plane due to a steep
trapping potential in the z-direction.  Hence one finds 
\begin{equation}
\mathbf{B}_{\mathrm{eff}} =  \hat{\mathbf{e}}_{z}\, \hbar k \frac{\partial
}{\partial x} \sin^2\theta.  
\label{B-eff-OAM-0}
\end{equation}
The field strength $B_{\mathrm{eff}}$ depends generally on the $x$ coordinate
and has a weak $y$-dependence as long as the paraxial approximation holds. 

If we are interested in fractional quantum Hall physics and thus in the
possibility to enter the lowest Landau level (LLL) regime we have to estimate
the maximum strength of the magnetic field. For this we determine the minimum
area needed for a magnetic flux corresponding to a single flux quantum
$2\pi\hbar$. From Eq.~(\ref{B-eff-OAM-0}) we recognize that this area is given
by the product $\lambda\, x_{\mathrm{eff}}$, where $x_{\mathrm{eff}}$ is the
effective separation between the two beam center.  To reach the LLL in a
two-dimensional gas the atomic density has thus to be smaller than one atom per
$\lambda\, x_{\mathrm{eff}}$. 

The above analysis holds as long as the atoms move sufficiently slow to remain
in their dark states.  This is the case if the adiabatic condition
\cite{juzeliunas05} holds: $\Omega\gg F$, where $F=
\left|\nabla\zeta\cdot\mathbf{v}\right|/(1+|\zeta |^2)$ reflects the two-photon
Doppler detuning.  In the present situation we have 
\begin{equation}
F^2 = \cos^2\theta \left[\left( v_{x}\frac{\partial }{\partial x }|\zeta
|\right) ^{2}+\left( |\zeta |kv_{y}\right) ^{2}\right]\ll\Omega^2.  
\label{F}
\end{equation}
The adiabatic condition implies that the rms Rabi frequency
$\Omega=(|\Omega_c|^2+|\Omega_p|^2)^{1/2}$ should be much larger than the time
an atom travels a characteristic length over which the amplitude or the phase of
the ratio $\zeta =\Omega _{p}/\Omega _{c}$ changes considerably.  For atoms
moving along the $y$ axis, such a length is $1/k\approx 1/2k_p\sim
10^{-7}\,\mathrm{m}$. On the other hand, the Rabi frequency can be of the order
of $10^7$  to $10^8\,\mathrm{s}^{-1}$ \cite{hau99}. Therefore, the adiabatic
condition should hold for atomic velocities up to  meters per second. 

The above estimation does not take into account a finite lifetime of the excited
atoms, typically $\tau_3\sim 10^{-7}\,\mathrm{s}$. If this is included, the
atomic dark state acquires a finite lifetime $\tau_D\sim\tau_3\Omega^2/F^2$ due
to nonadiabatic coupling \cite{juzeliunas05}: For instance, if the atomic
velocities are of the order of a centimeter per second, the atoms should survive
in their dark states up to a second. 

Much larger atomic velocities are possible however as long as the velocity
spread $\Delta \mathbf{v}$ is much smaller than the central velocity
$\mathbf{v}_{0}$.  For atoms moving along the  $y$ axis, one can set a
two-photon detuning $\omega_{21}=-(k_p+k_c)v_{0}$ to compensate the Doppler
shift associated with $\mathbf{v}_{0}$.  In that case it is the velocity spread
$\Delta \mathbf{v}$ rather than the whole atomic velocity $\mathbf{v}$ that
determines the non-adiabatic term $F$.  For instance, in a recent experiment
\cite{Stamper05prl} on propagation of a BEC in a waveguide, the central atomic
velocity is $5\,\mathrm{cm/s}$, whereas the velocity spread is only
$1.4\,\mathrm{mm/s}$. Note that the two-photon detuning will also lead to a
transversal slope in the trapping potential represented by the term with
$\omega_{21}$ in Eq.~(\ref{vext}).   

Let us assume that both the control and probe beams are characterized by
Gaussian profiles with the same amplitude $\Omega _{0}$ and width $\sigma $:
\begin{equation}
|\Omega _{j}|=\Omega _{0}\exp \left( -\frac{(x-x_{j})^{2}}{\sigma ^{2}}\right),
\qquad j=p, c.
\label{Omega-j}
\end{equation}
In the paraxial approximation, the Gaussian beams have the width $\sigma \equiv
\sigma (y)=\sigma_{0} [1+(\lambda y/\pi \sigma_{0} ^2)]^{1/2}$, where
$\sigma_{0}\equiv \sigma (0)$ is the beam waist and $\lambda$ is the wavelength.
Since $k_p\approx k_c\approx k/2$, we have  $\lambda\approx 4\pi/k$ both for the
control and probe beams.  We are interested mostly in distances $|y|$ much less
than the confocal parameter of the beams $b=2\pi \sigma_{0} ^2/\lambda \approx
k\sigma_{0} ^2/2$.  For such distances, $|y|\ll b$, the width  $\sigma (y)$ is
close to the beam waist:  $\sigma (y)\approx \sigma_{0}$.

Suppose the beams are centered at $x_{p}=x_{0}+\Delta x/2$ and
$x_{c}=x_{0}-\Delta x/2$, The intensity ratio reads then: $|\zeta |^{2}\equiv
|\Omega _{p}/\Omega _{c}|^2=\exp [(x-x_{0})/a]$, where $a\equiv a (y)=\sigma
^{2}/4\Delta x$ is the relative width of the two beams.  Thus we have 
\begin{eqnarray}
\mathbf{B}_{\mathrm{eff}} &=& -\hbar k\frac{1}{4a\cosh ^{2}\left(
(x-x_{0})/2a\right) }\mathbf{e}_{z},
\label{B_eff^D-gaus} \\
V_{\mathrm{eff}}(\mathbf{r})&=&V_{\mathrm{ext}}(\mathbf{r})+\frac{\hbar
^{2}k^{2}}{2m}\frac{\left( 1+1/4a^{2}k^{2}\right)} {4\cosh ^{2}\left(
(x-x_{0})/2a\right) }.
\label{V_eff^D-gaus}
\end{eqnarray}

It is evident that both $\mathbf{B}_{\mathrm{eff}}$ and
$V_{\mathrm{eff}}(\mathbf{r})$ are maximum at the central point $x=x_{0}$ and
decrease quadratically for $| x-x_0 | \ll a$.  Similar to
Ref.~\cite{juzeliunas05}, the term quadratic in the displacement $x-x_{0}$ can
be cancelled in the effective trapping potential (\ref{V_eff^D-gaus}) by taking
an external potential $V_{\mathrm{ext}}$ with the appropriate quadratic term. The
frequency of the external potential fulfilling such a condition is
\begin{equation}
\omega_{\mathrm{ext}}=\frac{\hbar k}{4am}\sqrt{1+1/4a^2 k^2}\, .
\label{omega-ext}
\end{equation}
With this the overall effective trapping potential becomes constant up to terms
of the fourth order in $x-x_0$.  In the vicinity of the central point ($| x-x_0
| \ll a$) the magnetic field strength is: $B_{\mathrm{eff}}\approx \hbar k/4a$.
The corresponding magnetic length and cyclotron frequency are: $\ell _{B}\approx
\sqrt{\hbar /B_{\mathrm{eff}}}= 2\sqrt{a/k}$ and $\omega _{c}=B/m\approx\hbar
k/4am$.  The magnetic length $\ell _{B}$ is much smaller than the relative width
of the two beams $\ell _{B} \ll a$ provided the latter is much larger than the
optical wave length: $ak\gg 1$.  In that case many magnetic lengths fit within
the interval $\left| x-x_{0}\right| < a$ across the beams. Furthermore the
cyclotron frequency equals then approximately to the frequency of the external
trap: $\omega_c \approx \omega_{\mathrm{ext}}$, both of them being much less
than the recoil frequency.

\begin{figure}
\begin{center}
\includegraphics[width=8.5cm]{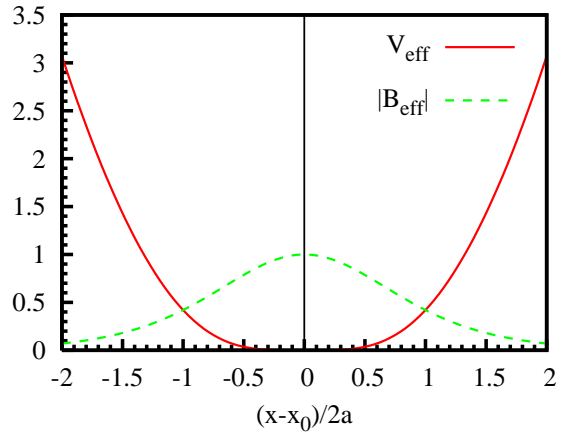}
\end{center}
\caption{(Color online) Effective trapping potential $V_{\mathrm{eff}}$ and
effective magnetic field $B_{\mathrm{eff}}$ produced by counter-propagating
Gaussian beams. The external harmonic potential $V_{\mathrm{ext}}$ cancels the
quadratic term in the overall potential $V_{\mathrm{eff}}$. The effective
magnetic field is plotted in the units of $B_{\mathrm{eff}}(0)\equiv \hbar
k/4a$, whereas the effective trapping potential is plotted in the units of
$\hbar \omega_{\mathrm{rec}}(1+1/4a^{2}k^{2})$, with
$\omega_{\mathrm{rec}}=\hbar k^{2}/2m$.}
\label{fig:2} 
\end{figure}

Figure \ref{fig:2} shows the effective trapping potential and effective magnetic
field calculated using Eqs.~(\ref{B_eff^D-gaus}) and (\ref{V_eff^D-gaus}), with
the external harmonic potential $V_{\mathrm{ext}}$ of frequency
$\omega_{\mathrm{ext}}$ (Eq.\ (\ref{omega-ext})) added to cancel the quadratic
term in the overall potential $V_{\mathrm{eff}}$.  The magnetic field is seen to
be close to its maximum value in the area of constant potential where
$|x-x_0|\ll a$.  For larger distances the effective trapping potential forms a
barrier, so the atoms can be trapped in the region of large magnetic field.

In summary, we have shown how to create an effective magnetic field in
ultra-cold gases with a planar geometry using two counter-propagating laser
beams acting on three-level atoms in the EIT configuration. If the amplitude
ratio of the two beams changes substantially in the transverse direction, an
effective magnetic field appears in the plane perpendicular to the propagation
direction of the beams.  This can be achieved if the the beams are shifted
relative to each other (see Fig.~\ref{fig:1}), such that they have a relative
OAM.

The suggested method provides a possibility to create an effective magnetic
field over an extended area along the propagation direction.  This allows for a
geometrical setup similar to that used in solid-state systems for classical and
quantum Hall measurements. In particular, finite currents perpendicular to the
magnetic field are possible and Hall ``voltages'' can be detected by observing
changes in the chemical potential perpendicular to both the current and magnetic
field. Finally the suggested method is much more robust than that of
Refs.~\cite{juzeliunas04,juzeliunas05}, as it does not require vortex light
beams.

This work was supported by the Marie-Curie Training-site at the Technical
University of Kaiserslautern, the Royal Society of Edinburgh, as well as the
Alexander von Humboldt Foundation through the collaborative grant between the TU
Kaiserslautern and the Institute of Theoretical Physics and Astronomy of Vilnius
University.

\end{document}